# Polarization-Controlled Coherent Phonon Generation in Acousto-Plasmonic Metasurfaces

Norberto D. Lanzillotti-Kimura[1,2,*], Kevin P. O'Brien[1,*], Junsuk Rho[1], Haim Suchowski[1], Xiaobo Yin[1], and Xiang Zhang[1,3]

[1]Nanoscale Science and Engineering Center, 3112 Etcheverry Hall,
University of California, Berkeley, California 94720-1740, USA

[2] Centre de Nanosciences et de Nanotechnologies, CNRS, Université Paris-Sud,
Université Paris-Saclay, C2N Marcoussis, 91460 Marcoussis, France

[3]Materials Sciences Division, Lawrence Berkeley National Laboratory,
1 Cyclotron Road, Berkeley, California 94720, USA

[*]Equally contributing authors

Acoustic vibrations at the nanoscale (GHz-THz frequencies) and their interactions with electrons, photons and other excitations are the heart of an emerging field in physics: nanophononics. The design of ultrahigh frequency acoustic-phonon transducers, with tunable frequency, and easy to integrate in complex systems is still an open and challenging problem for the development of acoustic nanoscopies and phonon lasers. Here we show how an optimized plasmonic metasurface can act as a high-frequency phonon transducer. We report pump-probe experiments in metasurfaces composed of an array of gold nanostructures, revealing that such arrays can act as efficient and tunable photon-phonon transducers, with a strong spectral dependence on the excitation rate and laser polarization. We anticipate our work to be the starting point for the engineering of phononic metasurfaces based on plasmonic nanostructures.

Coherent acoustic phonons can reach frequencies up to a few THz, with characteristic wavelengths in the nanometer scale, and are capable of interacting with photons, electrons, and other phonons. Based on these characteristics, several advances have been recently reported like phonon lasers (SASERs),[1,2] the study of novel physical phenomena, the emergence of high frequency cavity optomechanics,[3–5] and high frequency acousto-optic modulators to name a few. The generation and detection of high frequency acoustic phonons usually relies on the use of short laser pulses in a pump-probe type of experiments.[6] Typical transducers include metallic[7] and semiconductor thin-films[8,9], multilayers[10,11] and microcavities.[5,12–14] Great efforts have been made in the optimization of the light-matter interactions in these systems, and in the confinement and engineering of the acoustic and electric fields to improve the efficiency, tailor the spectrum and control the dynamics of the acoustic phonons.[15] However, one major challenge is still the development of a tunable source of high frequency acoustic phonons capable of replacing the standard metallic thin-films. Several approaches have been recently proposed to solve this problem using optimized multilayers,[16] optical microcavities,[13,17,18] metallic nanoparticles[19,20] and semiconductor nanostructures and quantum dots to name a few[8,21–23]. The use of localized surface plasmons is an alternative to overcome the aforementioned problem that has started to be exploited.[24–30] We propose the use of plasmonic metasurfaces as light-phonon transducers. By controlling the polarization we generate quasi-monochromatic acoustic phonons with different frequencies. Spatially localized surface plasmons show strong electronic resonances that allow their use for the design of optical nanoantennas and metamaterials, enabling new ways of capturing and controlling light.[31–34] These resonances can be strongly dependent on polarization, allowing a selective coupling and control of the light with the nanostructures.[35] A nanostructured surface or metasurface can be used to enhance light-matter interactions, to tailor wavefronts, and change phase profiles in laser beams.[36,37] Usually losses are the main limitation or drawback in the development of applications based on plasmonic resonances; in the study of acoustic phonon dynamics, on the contrary, all the desired effects such as generation of phonons, modification of the optical properties of the materials and good coupling with light, are strongly dependent on the coupling of the incident light, how much is absorbed, and how strongly the reflectivity changes in the surroundings of the used wavelengths. Similarly to what happens in optical microcavities, all these factors contribute positively to have enhanced signals in the generation of coherent phonons.[15] In this work we show how an optimized plasmonic metasurface can act as a high-frequency phonon transducer, where vibrational modes of different frequencies in the GHz range can be selectively excited and detected. We report pump-probe type coherent phonon generation experiments in metasurfaces composed of an array of gold nanobars. The antennas work for both the generation and the detection processes, enhancing the signals, and allowing polarization sensitive experiments. In this way, by combining different bar orientations and sizes on the metasurface we engineered a polarization-dependent coherent-phonon generation surface.

The polarization-dependent optical resonances and the phononic modes of the metallic nanobars forming the metasurfaces are determined by their geometry -size and shape- and material properties (index of refraction and sound velocity, respectively). We show how an array of metallic optical nanoantennas optimized to work at visible wavelengths can be tailored to generate and detect acoustic phonons of variable frequencies in the GHz range.

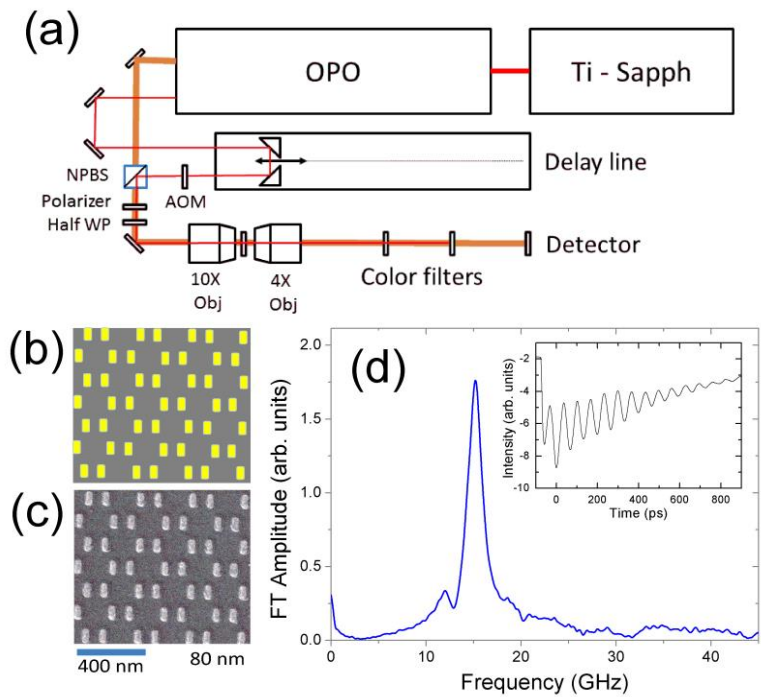

Figure 1. (a) Schematics of the experimental setup. (b) Schematics of a simple metasurface formed by parallel metallic bars. (c) Scanning electron micrography of the sample represented in b. (d) Coherent acoustic spectrum of the metasurface. The main peak corresponds to a longitudinal mode along the bars. The experimental time trace is shown in the inset.

To generate and detect ultrahigh frequency phonons we used a two-color pump-probe setup. A scheme of the experimental setup is depicted in Fig. 1a. A pump pulse excites the sample generating acoustic phonons by a thermoelastic effect; these phonons modulate the dielectric constant of the gold, and a second, delayed pulse detects the differential changes in the reflectivity due to the presence of the acoustic phonons. In Fig. 1b we show a scheme of a simple plasmonic metasurface formed by an array of parallel 90 nm gold bars. The scanning electron micrograph of the structure is shown in Fig. 1c. The phononic spectrum (Fig. 1d) presents a single peak at 15.2 GHz that corresponds to the longitudinal mode along the bars. The frequency is mainly determined by the length of the bars, and the full width at half maximum is due to non-homogeneities in the size and shape of the bars forming the metasurface. The measured time trace is shown in the inset, where clear oscillations can be observed.

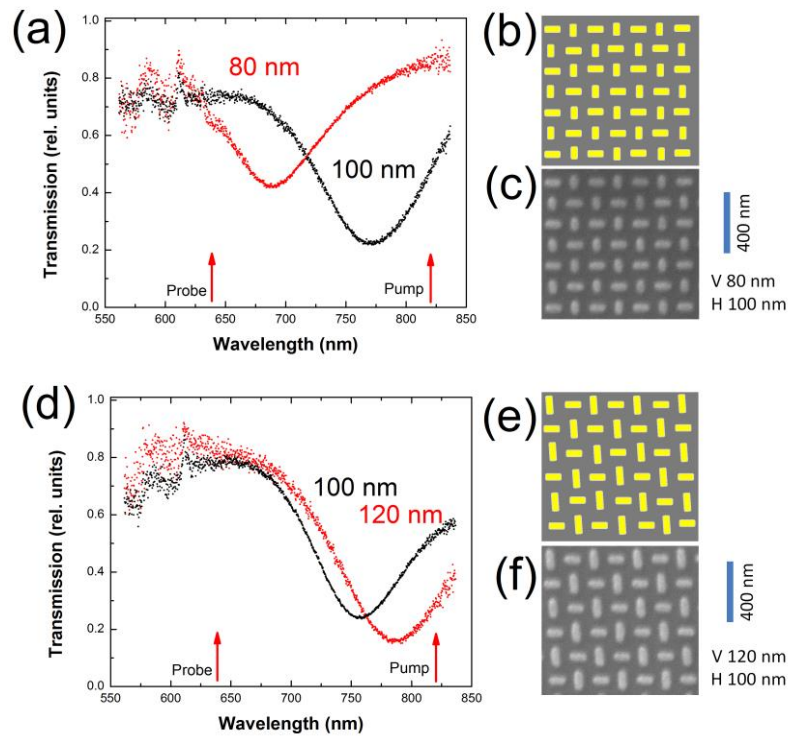

Figure 2. (a), (d) Measured optical transmission of the two reported metasurfaces. The transmission for the vertical (horizontal) polarization is plotted in red (black). (b), (e) Schematics of the studied plasmonic samples. (c), (f) Scanning electron micrographies of the metasurfaces. (a)- (c) correspond to the crossed bars metasurface with 100 and 120 nm length bars; (d)- (f) to crossed bars metasurface with 80 and 100 nm length bars, respectively.

The light-matter coupling can be controlled by means of plasmonic resonances. In this work we propose to use that coupling to control the transduction of GHz acoustic vibrations in nanostructures. We report the results on two different plasmonic metasurfaces. In Fig. 2 we show the optical transmission curves (panels a and d), the schematics of the surfaces (panels b and e) and the scanning electron micrographs (panels c and f) of the two samples. The samples are metasurfaces formed by crossed bars (Fig. 2b and 2e). They show an optical response that is strongly dependent on the polarization of the incident light. As it can be seen in the transmission curves presented in Figs. 2a and 2d, different polarizations will have associated different resonances due to the different sizes of the bars. This dual dependence of the resonance on the size and the orientation of the bars will be used to selective excite and detect tunable quasi-monochromatic phonons. The positioning of the bars was optimized to minimize the optical and acoustic cross-talking among different antennae.

In Fig. 3a (3b) we show the results of the pump-probe experiments for two different polarizations corresponding to the crossed bar sample A (B). In Fig. 3a each spectrum is characterized by a principal peak located at 12.9 and 15.5 GHz for the horizontal and vertical polarizations, respectively. The lowest frequency peak is associated to the longer bars (~100 nm) while the higher energy peak is originated in the shorter bars (~80 nm). The bars couple efficiently with the incident light only when the polarization is aligned with the main axis of the bars (both for the pump and probe pulses). In this way, each set of bars can act as a polarization sensitive transducer. In Fig. 3b we show similar results for sample B, in which the shorter bars (~100 nm) are in the horizontal orientation and the longer bars (~120 nm) are vertically oriented, generating the higher and lower frequency peaks respectively.

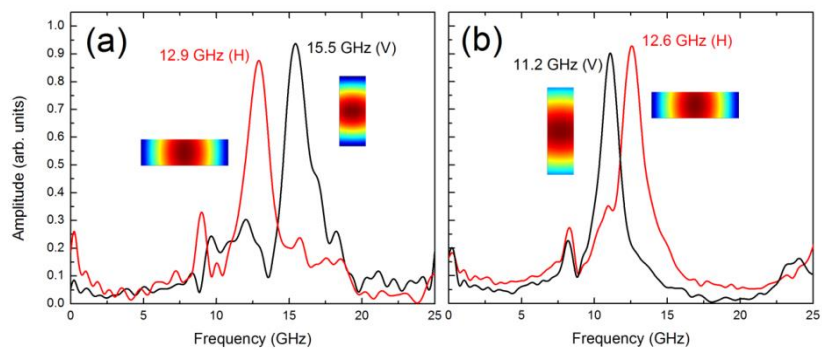

Figure 3. Polarization-controlled coherent phonon generation in the crossed-bars metasurfaces. Panel a (b) corresponds to the sample with bars of 80 and 100 nm (100 and 120 nm), for both vertical (black) and horizontal (red) polarizations. The black curves have been multiplied by a factor of 1.8 for clarity reasons. The simulated strain profiles of the modes are also presented.

In order to verify that the observed phonons actually correspond to the eigenmodes of the bars we calculated the eigenfrequencies of cylindrical tubes, which is a good representation of our nanostructures. Open cylindrical tubes resonate at the approximate frequencies $f=nv/[2(L+\alpha)]$, where n is a positive integer (1, 2, 3...) representing the resonance node, L is the length of the bar, v is the speed of sound in Au, and $\alpha$ is a constant linked to the transverse area of the bar. This equation compensates for the roughness, the energy leakage to the substrate, the dispersion in the size of the ensemble of bars, and the fact that the end facet of the bar is not the exact point at which a sound wave is reflected. In our analysis a sound velocity v=3240 m/s for Au was used, and the best fit is obtained for $\alpha$ = 26 nm. The parameter that determines the transmission of acoustic waves between the bars and the substrate is the acoustic impedance –product of the sound velocity and mass density. The high acoustic impedance mismatch between the Au bars and the glass substrate establish long lifetimes of the acoustic vibrations within the metallic structures, supporting the assumption that the FWHM is mainly due to the non-homogeneous distribution in sizes.

Using this equation we calculated the first mode of each structure. The results shown in Table 1, where a very good agreement is observed between the experimental results and the predicted eigenfrequencies.

| Nominal length | Experimental | Theory - Adjusted |
|---|---|---|
| 80 nm | 15.5 GHz | 15.3 GHz |
| 100 nm | 12.9 GHz | 12.9 GHz |
| 120 nm | 11.2 GHz | 11.1 GHz |

Table 1. Experimental and theoretical resonance frequency of the bars forming the crossed-bars metasurfaces. The experimental values correspond to the maximum intensity of the measured peak, the theoretical values were obtained using equation 1.

## Conclusions

We have experimentally demonstrated the transduction of different acoustic phonon frequencies using an array of crossed metallic bars, which was selectively excited/probed by changing the polarization of the excitation light. To take advantage of the vast possibilities of plasmonic nanostructures, and thus of plasmonic metasurfaces, more complex unit cells can be conceived. The presented results open the way to the development of acousto-optically functionalized surfaces capable of generating tailored vibrational frequencies in the 1 – 20 GHz range. This range is technologically relevant for deep profiling and nanostructures characterization using picosecond ultrasonic techniques. The reported results represent one of the first steps towards the engineering of complex acoustic phonons sources based on plasmonic nanostructures, replacing the standard metallic thin-films, and multilayers, and appearing as a strong alternative to other high frequency sources like nanoparticles and semiconductor quantum dots. In addition the presented crossed-bar nanostructured thin layer can be easily integrated in existent structures like planar optical microcavities and phonon nanocavities.

## Methods

Sample fabrication

The metasurfaces were fabricated on top of a quartz substrate via Electron-Beam Lithography (EBL), followed by a metal lift-off process. 2nm thick Indium-Tin-Oxide (ITO) is deposited on top of the quartz substrate by sputtering (Auto 306, Edwards) as EBL conductive layer purpose. After defining nanometer scale metasurface patterns in the bi-layer photoresist with 250 nm thick methyl methacrylate (MMA-EL8) and 50 nm thick polymethyl methacrylate (PMMA-A2) with high-resolution EBL (CABL-9000C, Crestec), electron beam evaporation system (Solution, CHA) is then used to deposit the chromium and gold thin films. Finally, sample is soaked in gently sonicating acetone to lift off the photoresist layer. The width of the bars was approximately 40nm, the thickness of the gold was 35nm. 2nm thin chromium layer is used to enhance adhesion between quartz surface and gold layer. In addition to the two studied samples, three control samples were fabricated. The control samples consist only of vertical bars, with sizes matching the ones used for the crossed bars samples.

Optical transmission experiments

To characterize the plasmonic response of the metasurfaces we performed optical transmission experiments (see Fig. 2). A super continuum light source was used and it was focalized using a microscope objective. We collected the light with a second microscope objective and dispersed the signal using a spectrometer. The signal was collected using a coupled charge device (CCD) camera. The signal was normalized by the transmitted light through the substrate.

Coherent phonon generation experiments

To characterize the samples we setup a two color transmission pump-probe experiment. Two collinear beams were focalized into the samples onto a ~5um spot using a 10X microscope objective. The signal was collected using a second microscope objective of larger numerical aperture (50X). We used Ti-Sapphire laser delivering ~100 fs pulses coupled with an Optical Parametric Oscillator (OPO) and a Second Harmonic Generator (SHG). Unless otherwise stated, we used a wavelength $\lambda$=820 nm (Ti-Sapph) for the pump and $\lambda$=630 nm (OPO+SHG) for the probe. We modulated the pump beam at a fixed frequency of 90 KHz using an acousto-optic device, in order to allow a synchronous detection with a lock-in amplifier. The signal from the probe was collected with a fast detector, after passing through a color filter to eliminate the signal contamination from the pump. Typical powers of 15 and 5 mW were used for the pump and probe beams, respectively. A mechanical stage was used to introduce a delay between the pump and probe pulses. These beams were originally cross-polarized and are recombined by a polarizing beam splitter. A polarizer and a half-waveplate were introduced in the beams path to have a variable and identical polarization in both beams. Time traces of 2000 ps were taken and Fourier transformed to analyze the frequency components of the phonon spectrum. All the experiments were performed at room temperature.

## Contributions

NDLK and KPO contributed equally to this paper. NDLK proposed the concept. JR fabricated the samples. NDLK and KPO set up and performed the experiments and analyzed the data. XZ guided experimental and theoretical investigations. All authors discussed the results and contributed with the writing of the paper.

## Acknowledgements

This work is supported by the Office of Naval Research (ONR) MURI program under Grant No. N00014-13-1-0631 and Samsung Electronics.